\documentstyle[epsf]{mn}

\title{ Microarcsecond instability of the celestial reference frame.}
\author[M. V. Sazhin, V.E. Zharov, A.V.Volynkin,  T.A.Kalinina ]
       {M. V. Sazhin, V.E. Zharov, A.V.Volynkin,  T.A.Kalinina \\
        Sternberg Astronomical Institute , Moscow 119899, Russia}
\date{Accepted .
      Received ;
      in original form }

\begin{document}

\maketitle

\begin{abstract}
The fluctuation of the angular positions of reference extragalactic radio
and optical sources under the influence of the irregular gravitational field
of visible Galactic stars is considered. It is shown that these angular
fluctuations range from a few up to hundreds of microarcseconds. This leads
to a small rotation of the celestial reference frame. The nondiagonal
coefficients of the rotation matrix are of the order of a microarcsecond.
The temporal variation of these coefficients due to the proper motion of
the foreground stars is of the order of one microsecond per 20 years.
Therefore, the celestial reference frame can be considered inertial
and homogeneous only to microarcsecond accuracy. Astrometric catalogues with
microarcsecond accuracy will be unstable, and must be reestablished every
20 years.
\end{abstract}


\section*{Introduction}

Modern Very Long Baseline Interferometry (VLBI) observations approach
microarcsecond accuracy in the determination of the positions of extragalactic
radio sources. This accuracy will soon reach the fundamental limit of
positional measurements associated with the curvature of space--time. Optical
astrometric experiments currently under consideration (the GAIA project) will
also be able to reach microarcsecond accuracy. Here, we consider
the effect of weak microlensing on the rotation of the radio celestial
reference frame based on radio quasars, though this effect is general, and
will also be relevant, in principle, to the corresponding optical reference
frame.

Photons of a radio or optical source (a quasar, for instance) move along a
path that is defined by the gravitational field of all the stars in our
Galaxy.  The stars in our Galaxy also move. As a result, their gravitational
field is nonstationary. Thus, the paths of photons are not straight lines,
but rather curves along nonstationary trajectories.

Two photons that leave the quasar at different moments of time move
along different paths.  Therefore, the directions in space from which an
observer detects these photons will be different. This causes fluctuations
in the position of the quasar. The aim of this paper is: 1) to calculate the
value of variations of the angular positions of selected quasars due to
this effect and 2) determine the influence of the gravitational field
of visible Galactic stars on the stability of the celestial reference
frame.

The difference between the real and apparent angular positions is proportional
to the mass of the body that bends the trajectory of the photon. The lower
the mass of a body, the less the bending of the trajectory of a test photon.
The main contribution to the effect, therefore, comes from the most massive
and most dense populations of stars.

Star-like objects can act on the propagation of light in two ways. The
first is via microlensing, which forms two images (in place of the single
true image) separated by an angular distance of the order of $10^{-3}\; {\rm
arcsec}$. A close angular coincidence of the background quasar and deflecting
body is necessary in this case. This microlensing effect can act as a space
telescope with extremely high resolution \cite{b4}. The microlensing
effect was predicted by Paczinsky \cite{pac86} and later discovered by the
MACHO, EROS, and OGLE groups (Alcock et al. 1993; Aubourg et al. 1993;
Udalski et al. 1994).

As an ordinary lens that has an infinite radius of gravitational action,
a foreground object can change the trajectory of photons from a background
object even if the microlensing effect is absent.  Therefore, the second way
star-like objects can act on the propagation of light is by producing a
small deviation of a photon trajectory from a straight line when the
deflecting body is relatively far from the unperturbed star--observer
trajectory. A body with mass $M$ and impact parameter $p$ relative to the
unperturbed trajectory bends a light beam by an angular amount $2M/p$.  The
observer measures an angular deviation $M/p$ from the unperturbed position
of the quasar. This idea was considered in Sazhin (1996) in connection
with the limit of astrometric positional accuracy and more
theoretically in Zhdanov (1995).  Let us consider a population with a
local density (in the vicinity of the Sun) of the order of $n=0.1\;
{\rm pc}^{-3}$ with masses for individual objects of the order of
$0.1\, M_{\odot}$. If an extragalactic object is observed, one can
expect that at least one deflecting object will have an impact
parameter of the order of
$$
p=(\pi R_H\, n )^{-\frac{1}{2}}.
$$
Here, $R_H$ is the radius of the Galactic halo. In this case, the
observer expects to see an angular deviation of the apparent position
of the order of   $1\ {\rm \mu arcsec=\mu as}$ from the true position
of this extragalactic object.

Only the second propagation effect noted above will be considered here.
The first effect (microlensing) is a very rare event:  it is necessary
to observe one or two million stars in order to detect one
microlensing effect. On the contrary, the second effect will act to
some extent on every celestial source of light. One can expect a
deviation of the order of $1\;{\rm \mu as}$ for every extragalactic
source. Thus, the angular range of interest to us is from 1 ${\rm \mu
as}$ to 1000 ${\rm \mu as}$, which is well below the level of the
microlensing effect, but is measurable using modern VLBI techniques.
The effect of weak microlensing can already be probed at
radio wavelengths due to the high accuracy of VLBI
observations.

\section*{Weak microlensing}

Let us consider a situation in which there are an extragalactic source
(source of a test photon S), a massive body B with mass $M_B$ (source
of a gravitational field that bends the trajectory of the test
photon), and an observer who detects the test photon (bent by the
gravitational field of B). The direction of the incident photon does
not coincide with the straight line connecting the extragalactic
source and the observer.  Therefore, the apparent celestial position
of the extragalactic source does not coincide with its true
position.

The effect is defined by the Einstein cone, which has the value
$$
\varphi^2_E=\frac{4GM_B}{c^2}\frac{L_{SB}}{L_{OB}(L_{SB}
+L_{OB})}
$$
where $L_{OB}$ and $L_{SB}$ are the distances from light
source S to the body B and from the observer O to the body B,
respectively. In the microlensing effect, two images appear, which
have angular distances from the body B:
$$
\varphi_1=\frac{\varphi}{2}+\frac{1}{2}\sqrt{\varphi^2 +4
\varphi^2_E}
$$
and
$$
\varphi_2=\frac{\varphi}{2} -\frac{1}{2}\sqrt{\varphi^2 +4 \varphi^2_E}
$$
where $\varphi$ is the angular distance from B to the true position of
the quasar, and $\varphi_1$ and $\varphi_2$ are the angular distances
of the two images relative to B. The brightness of the second image
$\varphi_2$ is inversely proportional to the fourth power of
$\varphi$, and in the case $\varphi>>\varphi_E$, the observer cannot
see the second image due to its weakness. In this case, the first
image has approximately the same brightness as the intrinsic
brightness, without the microlensing effect. The separation of the
first image from the real position of the star is
$$
\delta \varphi=\frac{1}{2}\sqrt{\varphi^2 +4 \varphi^2_E}
-\frac{\varphi}{2} \sim \frac{\varphi_E^2}{\varphi}.
$$

The Galactic body B moves with some angular velocity $\Omega$ due to its
peculiar motion in the Galaxy. In this case, one can write an equation for
$\varphi(t)$ as a function of time:
$$
\varphi^2(t)=\varphi^2_p +\Omega^2 t^2
$$
where $\varphi_p$ is the angular impact parameter of the body relative to
the background quasar, and we take $t=0$ to be the moment of closest
approach. It is easy to recalculate this equation from the modulus of
the angular separation to the spherical coordinates $\alpha$ and
$\delta$.

\section*{Variations of the coordinates of reference quasars}

The question of the variations of the coordinates of selected quasars
is especially important because the 23rd General Assembly of the
International Astronomical Union decided that, as of 1 January, 1998,
the IAU celestial reference system is the International Celestial
Reference System, in replacement of the J2000 system realized by
the FK5 \cite{fri88}.  The ICRS is realized by the International
Celestial Reference Frame (ICRF), defined by the equatorial
coordinates of a set of selected extragalactic compact radio sources
determined using VLBI. The ICRF consists of a catalogue of equatorial
coordinates of these radio sources \cite{iers}.

To test the effect of gravitational refraction (or microlensing
effect), we used the catalogue published in the 1995 International
Earth Rotation Service (IERS) Annual Report
(RSC(IERS)95~C02)~\cite{z1}.  This includes a total of 607 objects
spread over the sky from declination $-85^\circ$ to $+85^\circ$.  The
uncertainties in the coordinates are from $50\div 2000\; {\rm \mu as}$.
There are 236 primary objects that are the most compact and best
observed; 321 secondary compact sources that may have very precise
coordinates in the future, when more observations are accumulated; and
50 sources that are complementary objects observed for optical frame
ties or other objectives.

The stability of the frame is based on the assumption that the sources
have no proper motion. The hypothesis that the sources are fixed is
used in performing the coordinate transformation between the celestial
and terrestrial reference systems. In reality, some of these sources
have significant structure on mas scales. Changes in the source's
brightness distribution can shift the effective brightness centroid of
the source and thus its coordinates. A small number of objects from the
reference frame exhibit such changes \cite{z2}, and they can reach a
few tenths of mas.

Gravitational refraction leads to changes in the coordinates
of {\it all} sources with time, and to a small rotation of the frame.
Here, we do not calculate the effect of gravitational refraction from
dark deflecting bodies and faint stars, but only from visible stars.  The
influence of dark bodies on the stochastic motion of apparent
extragalactic source positions was considered in \cite{b5}, and is of
the order of 1 $\mu$as. The rotation of the celestial reference frame
due to the influence of dark bodies and fainter stars will be
considered in a later study.

We used the Guide Star Catalogue (Lasker et al. 1990; Russel et al.
1990; Jenkner et al. 1990) and the HIPPARCOS catalogue \cite{z6} to
find stars whose angular distances from the extragalactic reference
sources do not exceed $2'$. A total of 170 quasars have a total of 313
nearby stars.  73 quasars have 86 neighboring stars in a circle with
radius $1'$. For most of these stars, we know neither their mass nor
their distance.  Two bright stars were found in the HIPPARCOS
catalogue. The proper motion, trigonometrical parallax, and spectral
type of both stars are known, so that it is possible to determine their
masses and distances.

To estimate the rotation of the true ICRF relative to the observed ICRF,
we used only the primary reference sources. 27 sources have nearby
stars, and 5 have two stars within $1'$.  The rotation vector ${\bf
\Theta}=(\theta_1, \theta_2, \theta_3)$ is defined by the
matrix
\begin{equation}
{\bf r_i}'= \left(\matrix {1       & -\theta_3& \theta_2 \cr \theta_3
              &  1        & -\theta_1\cr -\theta_2 & \theta_1  &   1
\cr} \right){\bf r_i},
\end{equation}
where ${\bf r_i}$ and ${\bf r_i}'$ are the radius vector of the $i$th
source before and after the deflection of the light from this source.
We estimated the distance to the star $L_{OB}$ using the
mass--luminosity relation \cite{z7}:
$$
{\rm lg}\frac{L_s}{L_\odot}=3.5{\rm lg}\frac{M_s}{M_\odot},
$$
where $L_s,L_\odot$ are the luminosities and $M_s,M_\odot$ the masses
of the star and the Sun, respectively. The mass of each star was
determined in the range from $ M_\odot$ to $30M_\odot$ using the
Salpeter mass function $dn/dM\sim M^{-2.35}$.  The distance $L_{OB}$
can be found from the equation
$$
{\rm lg}L_{OB}+0.2AL_{OB}=0.2(m+0.26)+1.75{\rm
lg}\frac{M_s}{M_\odot}.
$$
The apparent magnitudes $m$ of the stars lie in the range from
$11.3-15.8$.  The distances vary from 100 to 5500 pc. We assume here
that the total absorption $A$ in the vicinity of the Sun is $A=1.9^m\
{\rm kpc}^{-1}$ \cite{z8}.

We find that  ${\bf \Theta}=(-0.08\pm 0.02,-0.47\pm 0.03, -0.36\pm
0.03)\; {\rm \mu as}$ for various realizations of the Salpeter mass
distribution (the variances of ${\bf \Theta}$ correspond to the ranges
for various mass distributions). The change of ${\bf \Theta}$ with time
was estimated using a uniform distribution of possible proper motions
of the stars in the range $-100 \div 100\ {\rm \mu as/year}$.  The
value of the time derivative ${\bf \Theta}$ (or the angular rotation
rate of the ICRF) is of the order of $1\ {\rm \mu as}$ per 20 years,
but can reach $1\ {\rm \mu as}$ per 1 year. The number of primary
reference sources in the catalogue is 236. The resulting value of
${\bf \Theta}$ is defined by the $\sim 10\%$
of the sources that have nearby stars. The stochastic process of ray
deflection has an unknown distribution. If we increase the number of
primary reference sources, the value of ${\bf \Theta}$ will be
decreased, but the law by which ${\bf \Theta}$ diminishes is unknown.
Since this law has a nongaussian distribution, the diminishing of ${\bf
\Theta}$ will not be simply proportional to the inverse square root of
the size of the catalogue.

\begin{table*}
\caption{ Coordinates of quasars and nearby stars and the effect of
gravitational refraction. $\Delta\alpha$, $\Delta\delta$
are measured in ${\rm \mu as}$}.
\begin{center}
\begin{tabular}{|c|ccl|ccl|c|c|}
Source/Magnitude/ & \multicolumn{3}{|c|}{Right ascension}
&\multicolumn{3}{|c|}{Declination} &\multicolumn{2}{|c|}{Uncertainty}\\
Distance ('')   & h& m& s& $^\circ$& '& ''& s & ''\\
            &  &   &           &   &   &         &         &        \\
    0007+106& 0& 10& 31.005871 & 10& 58& 29.50408& 0.000018& 0.00042\\
      14.7 & 0& 10& 31.01     & 10& 58& 29.9    &         & \\
    0.358   &  &   & $\Delta\alpha=39.8$ & & & $\Delta\delta=219.6$ & & \\
            &  &   &           &   &   &         &         &        \\
    0111+021& 1& 13& 43.144949&   2& 22& 17.31639&  0.000014& 0.00038\\
        13.8& 1& 13& 43.13    &   2& 22& 17.8    &          & \\
    0.509   &  &   & $\Delta\alpha=-82.4$ & & & $\Delta\delta=133.7$ & & \\
            &  &   &           &   &   &         &         &        \\
    0735+178& 7& 38&  7.393743&  17& 42& 18.99868&  0.000003& 0.00005\\
         4.3& 7& 38&  7.37    &  17& 42& 19.0    &          & \\
    0.284   &  &   & $\Delta\alpha=-293.6$ & & & $\Delta\delta=-30.5$ & & \\
            &  &   &           &   &   &         &         &        \\
   0912+297 & 9& 15& 52.401619&  29& 33& 24.04293&  0.000017& 0.00034\\
       15.3 & 9& 15& 52.40    &  29& 33& 23.5    &          & \\
    0.496   & &   & $\Delta\alpha= 18.1$ & & & $\Delta\delta=-160.5$ & & \\
            &  &   &           &   &   &         &         &        \\
   1101+384 &11&  4& 27.313911&  38& 12& 31.79962&  0.000026& 0.00038\\
       12.8 &11&  4& 27.31    &  38& 12& 31.8    &          & \\
    0.054   &  &   & $\Delta\alpha=776.7$ & & & $\Delta\delta=1329.0$ & & \\
            &  &   &           &   &   &         &         &        \\
   1302-102 &13&  5& 33.015008& -10& 33& 19.42722&  0.000018& 0.00021\\
       14.8 &13&  5& 32.98    & -10& 33& 20.1    &          & \\
    0.812   &  &   & $\Delta\alpha=-62.6$ & & & $\Delta\delta=-77.0 $ & & \\
            &  &   &           &   &   &         &         &        \\
  1514-241  &15& 17& 41.813132& -24& 22& 19.47552&  0.000019& 0.00031\\
      13.3  &15& 17& 41.84    & -24& 22& 19.8    &          & \\
    0.536   &  &   & $\Delta\alpha=120.8$ & & & $\Delta\delta=-100.9$ & & \\
\end{tabular}
\end{center}
\end{table*}                          \

The largest effects of gravitational refraction were displayed by
several of the extragalactic sources (Table 1). The first row of Table
1 contains the name of the source, its right ascension and declination,
and their uncertainties as reported by the IERS. The second row shows
the magnitude and coordinates of the nearby star. In the third row, the
angular distance (in arcsec) and corrections for the quasar's
coordinates (in ${\rm \mu as}$) are shown. These coordinate corrections
were calculated assuming that the mass and distance of each star are
$M_s=5M_\odot$ and $L_{OB}=500$ pc.

\begin{table*}
\caption{ The coordinates of the quasars and HIPPARCOS stars }
\begin{center}
\begin{tabular}{|c|ccl|ccl|c|c|}
 Source/Number & \multicolumn{3}{|c|}{Right ascension}
               &\multicolumn{3}{|c|}{Declination}
               &\multicolumn{2}{|c|}{Uncertainty}\\
            & h& m& s& $^\circ$& '& ''& s & ''\\
            &  &   &           &   &   &         &         &        \\
    0459-753&  4& 58& 17.945614& -75& 16& 37.95439  & 0.000923& 0.00315\\
     23106  &  4& 58& 17.95    & -75& 16& 38.0      & 0.000037& 0.00060\\
            &  &   &           &   &   &         &         &        \\
    1213-172& 12& 15& 46.751743& -17& 31& 45.40314  & 0.000043& 0.00041\\
     59803  & 12& 15& 48.47    & -17& 32& 31.1      & 0.000042& 0.00049\\
\end{tabular}
\end{center}
\end{table*}

The exact effect of gravitational refraction can be
calculated for the two stars from the HIPPARCOS catalogue (Table 2).
The change of the stars' coordinates due to their proper motion leads
to a change of the angular distance to the nearby quasar. As a
result, a secular variation of the coordinates of the quasar will be
observed.  Figure~1 shows the change of the coordinates of the quasar
0459-753 over 100 years, from 1952 to 2050. During the period from
1986--1996 (when the observations were carried out) the right ascension
and declination of the quasar changed by up to 7.5 mas and 1.5 mas,
respectively.  The first star (with number 23106) has spectral type
$K1\,{\rm III}p$.  It is a star with mass approximately $4M_\odot$
\cite{z7}.  The parallax of this star is $3.43\pm 0.61$ mas and its
proper motion is $\mu_{\alpha}=-4.00\pm 0.57$ mas/year,
$\mu_{\delta}=-2.55\pm 0.78$ mas/year.

We suggest that the large uncertainties of the coordinates reported by
IERS (Table 2) could be connected with the apparent motion of the
quasar due to gravitational refraction. The calculated values of
$\Delta \alpha$ and $\Delta \delta$ are less than the observed values.
The path of the quasar (Fig.~1) depends on the star's parameters,
including its coordinates and mass.  If the mass is underestimated,
the true values of $\Delta \alpha$ and $\Delta \delta$ should be
larger. In the new version of the ICRF (available at IERS under the
label RSC(WGRF)95 R 01), the quasar 0459-753 is omitted (perhaps due
to large coordinate uncertainties due to the unidentified effect of
gravitational refraction).

Figure~2 shows the change of the
coordinates of the quasar 1213-172 over 100 years.  The second star
(with number 59803) has spectral type $B8\,{\rm III}$. It is a normal
giant with mass approximately $10M_\odot$ \cite{z7}. The parallax of
this star is $19.78\pm 0.81$ mas and its proper motion is
$\mu_{\alpha}=-159.58\pm 0.66$ mas/year, $\mu_{\delta}=22.31\pm 0.54$
mas/year. The effect of refraction is smaller in this case, though the
star is very close to the Sun (50 pc).

\begin{figure}
\epsfxsize=\columnwidth
\epsfbox{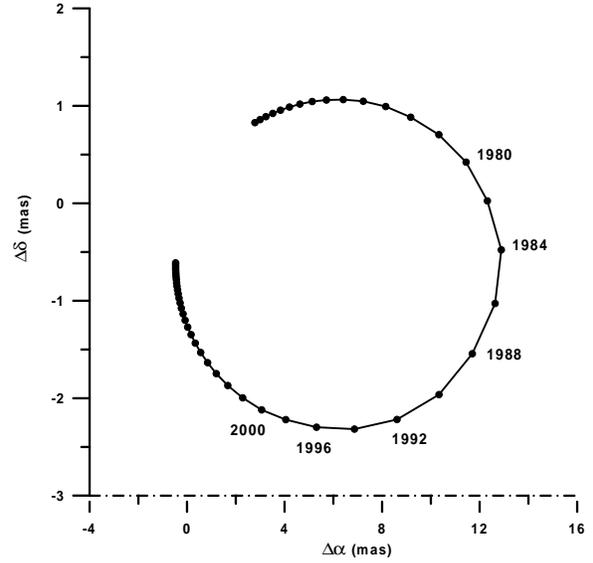}
\caption{The variation of the coordinates of the quasar 0459-753 due to
the proper motion of the star 23106.}
\end{figure}

\begin{figure}
\epsfxsize=\columnwidth
\epsfbox{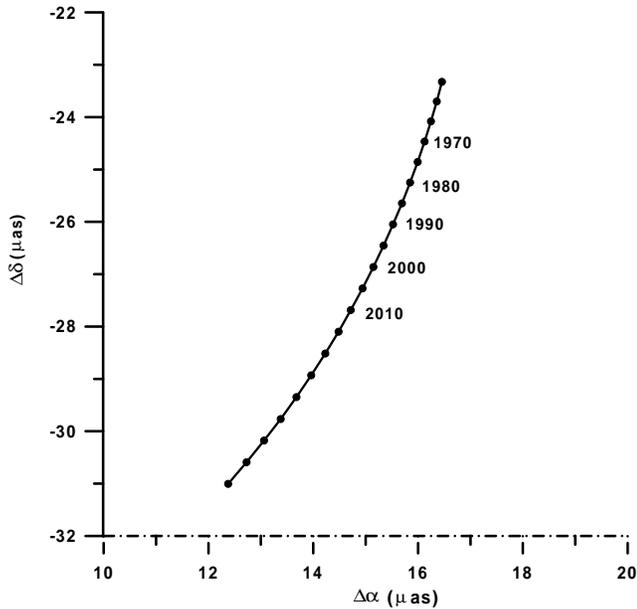}
\caption{The variation of the coordinates of the quasar 1213-172 due to
the proper motion of the star 59803.}
\end{figure}

\section*{Conclusion}

There are two important conclusions from our investigation.  First, we
have shown that variations of the angular positions of
reference quasars range from a few microarcseconds up to hundreds of
microarcseconds. The effect of weak microlensing leads to a small
rotation of the celestial reference frame.  The nondiagonal
coefficients of the rotation matrix are of the order of
microarcsecond.  The temporal variation of these coefficients due to
stars' proper motions is of the order of one microarcsecond per 20
years.  Because the proper motions of the majority of Galactic stars
are unknown, the extragalactic reference frame can be considered
inertial and homogeneous only to microarcsecond accuracy. The effect
considered here should be investigated using current VLBI catalogues;
however, this effect is general, and astrometric optical catalogues
with microarcsecond accuracy will also be unstable, and must be
reestablished about every 20 years. It is possible that one example of
weak microlensing has been found: the source 0459-753, which has a
nearby HIPPARCOS star and shows large uncertainties in its
coordinates.

\subsection*{Acknowledgments}

We thank Dr. A. Kuzmin, V. Sementzov, K. Kuimov, and M. Prokhorov for
a critical reading of the original version of the paper and
many helpful suggestions. We would also like to acknowledge an
anonymous referee whose comments and remarks substantially improved
our paper. This work was supported in part by the ``Cosmion'' Center
for Cosmo-Particle Physics, the Russian Foundation for Basic
Research (grants NN 97-02-17434, 97-05-64342, 98-05-64797), and
Russian Federal Programme "Integration" (project K0641).

\end{document}